\UseRawInputEncoding
\documentclass[twocolumn,           % Format : preprint, twocolumn
               showpacs,            % Pacs : showpacs, noshowpacs
               nopreprintnumbers,     % Preprint: preprintnumbers,
               aps,                 % Society: ...
               prd,          	    % Journal Style : pra, prb, prc, prd, pre,
               letterpaper,             % Size : a4paper, ...
              groupeaddress,      % Affiliation (Title) : groupedaddress,
               nofootinbib,         % Footnote: footinbib, nofootinbib
               tightenlines,        % Remove additional spaces in a line
               floats,floatfix,      % Floating pictures and tables
               showkeys
               ]{revtex4-1}
               
\usepackage[toc,page]{appendix}
\usepackage{graphicx}% Include figure files
\usepackage{dcolumn}% Align table columns on decimal point
\usepackage{bm}% bold math
\usepackage{amsmath}
\usepackage{amsfonts,amssymb}
\usepackage{soul}
\usepackage{color}
\definecolor{v}{rgb}{0.6, 0.2, 0.8} %comentarios VM
\definecolor{MAGA}{rgb}{0.1, 0.43, 0.75}
\definecolor{jm}{rgb}{0.13, 0.48, 0.64}

\usepackage{xcolor}
\usepackage{enumerate}
\usepackage{float}
\usepackage{subfigure}
\usepackage{multirow,tabularx, booktabs}
\usepackage{orcidlink}

\usepackage{silence}
\newtheorem{de}{Proof}[section] 
\newtheorem{lem}{Lemma}[section]

\begin{document}

\title{DESI and fast radio burst used to constrain modified theories of gravity}

\newcommand{\orcidauthorA}{0000-0003-0405-9344} % Alberto
\newcommand{\orcidauthorB}{0000-0002-6356-8870} % Miguel
\newcommand{\orcidauthorC}{0000-0001-9310-2935}
\newcommand{\orcidauthorD}{0000-0003-4446-7465} % Veronica

\author{J. A. Astorga-Moreno$^1$\orcidlink{\orcidauthorC}}
\email{jesus.astorga@cinvestav.mx, corresponding author}

\author{Miguel A. Garc\'ia-Aspeitia$^2$\orcidlink{\orcidauthorB}}
\email{angel.garcia@ibero.mx}

\author{A.  Hern\'andez-Almada$^3$\orcidlink{\orcidauthorA}}
\email{ahalmada@uaq.mx}

\author{V. Motta$^4$\orcidlink{\orcidauthorD}}
\email{veronica.motta@uv.cl}

\affiliation{$^1$Departamento de F\'isica, Centro de Investigaci\'on y Estudios Avanzados del IPN, Apartado Postal 14-740, 07000, CDMX, Mexico}

\affiliation{$^2$ Depto. de F\'isica y Matem\'aticas, Universidad Iberoamericana Ciudad de M\'exico, Prolongaci\'on Paseo \\ de la Reforma 880, M\'exico D. F. 01219, M\'exico}

\affiliation{$^3$ Facultad de Ingenier\'ia, Universidad Aut\'onoma de
Quer\'etaro, Centro Universitario Cerro de las Campanas, 76010, Santiago de 
Quer\'etaro, M\'exico}

\affiliation{$^4$Instituto de F\'isica y Astronom\'ia, Universidad de Valpara\'iso, Avda. Gran Breta\~na 1111, Valpara\'iso, Chile.}

%-------------------------------------------------------------------------------------------------
%-------------------------------------------------------------------------------------------------
\begin{abstract}
This paper is devoted to the study of the viability of DESI and the fast radio burst to constrain the free parameters of modified theories of gravity. Thus, we present a model supported in $f(R)$ gravity involving a function of the Ricci scalar named Starobinsky-type with the peculiarity that the non-commutative essence is intrinsic to the coefficients. Additionally, to understand the dynamics within a flat Friedmann-Lemaitre-Robertson-Walker universe, we explore the possibility of deriving a Friedman equation (in measure) that results from an adequate mathematical treatment. As we mentioned previously, to test the outlined model, a Monte Carlo Markov chain analysis is implemented, using cosmic chronometers, type Ia supernovae, Hydrogen II galaxies, Intermediate-luminosity quasars, Baryon Acoustic oscillations and Fast Radio Bursts data, to constraint the free parameters of the model and presenting $H(z)$, $q(z)$ and $\omega_{eff}(z)$. The final results are compared with the $\Lambda$CDM model and a robust discussion is presented about the viability of DESI and fast radio burst to constraint free parameters in specific to a modified theory of gravity.
\end{abstract} 
%\draft
%\date{=day}
\maketitle

%%%%%%%%%%%%%%%%%%%%%%%%%%%%%%%%%%%%
\section{Introduction}  \label{intro} 
%%%%%%%%%%%%%%%%%%%%%%%%%%%%%%%%%%%%

General Relativity (GR) is the cornerstone of the Modern Cosmology, being the Cosmological Standard model ($\Lambda$CDM model) constructed in the light of this outstanding theory. Despite the unprecedented success of this model - already confirmed by Cosmic Microwave Radiation (CMB) \cite{Planck:2018}, the Big Bang Nucleosynthesis (BBN) \cite{bbn}, Supernovae Type Ia (SNIa) \cite{Riess_2022, Perlmutter:1999}, among other observations - challenges such as the phenomenological understanding of the Dark Matter (DM) and Dark Energy (DE) are still unknown, even though they comprise almost the totality of our Universe. Additionally, issues like the $H_0$ tension, the CMB anisotropy anomalies, the impossibility to obtain the observed density energy for the Cosmological Constant (CC) from quantum vacuum fluctuations - showing a discrepancy of 120 orders of magnitude (see \cite{Alcantara-Perez:2023jbv} as a form to alleviate this problem) - have forced the community to explore modifications of the GR, mainly reflected in emergent DE models \cite{thomas, 
li,rast}. For example, focusing on the late-time accelerated expansion, some models suggest a change in the curvature by introducing a parameter that reproduces to make the transition softly for the Hubble parameter behavior \cite{curv}. Among the emergent DE models reported, some include different fluids derived from Quantum Field Theory, and their associated particles; others introduce modifications to RG, ignoring the necessity to add extra fluids to generate Universe acceleration. A broad discussion of these exotic proposals is found in \cite{tax}, where the increase of observational samples and the improvement of the instrumentation have led to a way to measure their real efficacy, allowing us to distinguish between them or provide a
viable alternative for the Standard Model of Cosmology. 

 In this vein, the modified theories of gravity that have received more attention than others are the Scalar Tension Theories,
notably including, in a historical and relevant order, the Kaluza-Klein theory, which introduces a four-dimensional spacetime with an extra dimension appearing as a spacetime scalar,  and the Jordan-Brans-Dicke formalism, where a scalar field (or better known as inflaton) plays the role of a time-varying gravitational constant \cite{brans}. 
 This hypothetical object drives to inflationary expansion, renewing interest in this kind of theories, whose key mission is to generalize the $\Lambda$CDM model via the scalar field as the dynamic dark energy component responsible for the accelerated expansion. Additionally, there is extensive literature on scalar field dark energy models \cite{quinta,kessen,quinto}, like quintessence or canonical scalar field models, kinetic dominated scalar field model (called k-essence), and the quintom model, in which the equation of state evolves across the cosmological constant boundary. Among the main formalisms, we have to mention Unimodular Gravity (UG) \cite{gao,uni-mig}, whose most prominent feature is the appearance of the CC in the respective equations of motions as a constant of integration, allowing its value to be fixed from the beginning. One of the simplest extensions of GR, usually called $f(R)$ gravity \cite{amen}, can be obtained by replacing $R$ in the Einstein-Hilbert action with a function that depends on the Ricci scalar. The first success for this model was the inflationary scenario introduced by Starobinsky \cite{star}, where $f(R)=R+\alpha R^2$ for $\alpha>0$, offering an exact de Sitter expansion when $R^2$ dominates, and leading to an exit from inflation if $R$ is dominant, \cite{k} 
 has shown that the model proposed by Starobinsky is consistent with the Planck-2018 data \cite{Planck:2018}.

Essentially, in the analysis shown below, a quadratic function of the Ricci scalar $f_{nc}$ is proposed, where the coefficients contain the parameter of non-commutativity. We point out
the importance of this parameter, as non-commutative cosmology has been extensively studied by many authors from both classical and quantum perspectives \cite{barb} with the aim of clarifying the role of non-commutativity in various aspects of the cosmological scenario. For our purposes, the outstanding contexts we can mention are non-commutative gravity, quantum, and inflationary cosmology \cite{infl,obr,infl2,grav}. It is important to highlight that the last two are particularly significant, as in the very early universe - when it was small and hot - the non-commutativity could
have played a relevant role in its evolution \cite{stoch}.

The outline of the paper is as follows. In Sect. \ref{f_nc.cosmology} we present the cosmology in $f_{nc}(R)$ gravity aimed at exposing the Friedmann equation in this approach, Sect. \ref{Datasets} is dedicated to show the datasets used to constrain the free parameters of the model, additionally in Sec. \ref{Results} we describe the main results of the paper and the ideas related
to $f_{nc}(R)$ gravity. Finally, Sect. \ref{Conclusions} is dedicated to conclusions and perspectives. We henceforth use units in which $\hbar=c=k_B=1$.

%%%%%%%%%%%%%%%%%%%%%%%%%%%%%%%%%%%%%
\section{$f_{nc}(R)$ Cosmology} \label{f_nc.cosmology}
%%%%%%%%%%%%%%%%%%%%%%%%%%%%%%%%%%%%%

By introducing a perfect fluid in the Energy-Momentum Tensor in the field equations \eqref{f.e}, we get the following Friedmann-type equations within a framework that considers the Ricci scalar function \eqref{fun.frncc} involving the parameter of non-commutativity

\begin{equation}
H^2=\frac{\Big[2\kappa\rho_i+a_{\eta,0}+a_{\eta,1}R-a_{\eta,2}R^2-6a_{\eta,2}H\dot{R}\Big]}{6(a_{\eta,1}+2a_{\eta,2}R)},\nonumber
\end{equation}
\begin{eqnarray}
2\dot{H}+3H^2=-\Big[2\kappa \omega_i \rho_i +a_{\eta,0}+a_{\eta,1}R+
a_{\eta,2}R^2 +\nonumber\\
2a_{\eta,2}\dot{R}+4a_{\eta,2}H\ddot{R}\Big]
\Big[2(a_{\eta,1}+2a_{\eta,2}R))\Big]^{-1},\label{h2}
\end{eqnarray}
being $H$ the Hubble parameter, $a_{\eta,j}$ the coefficients of $f_{nc}$, $\kappa=8\pi G_{N}$, with $G_N$
the Newtonian Gravitational constant, $\rho_i$ the densities of the cosmological fluids and $\omega_i$ their respective parameters in the EoS. The continuity equation is given by (see Appendix \ref{ap.b})
\begin{align}
\kappa\dot{\rho_i}+3H\Big[\kappa(\omega_i+1)\rho_i+f^{\prime}_{nc}(\bar{\gamma_\eta})\Big] &=0,
\end{align}
as an immediate consequence, we get for the usual cosmological fluids\footnote{This means matter and radiation.}
\begin{align}
\rho_i&=\frac{\rho_{0i}}{\kappa(\omega_i+1)}a^{-3(\omega_i+1)}-\frac{f^{\prime}_{nc}(\bar{\gamma_\eta})}{\kappa(\omega_i+1)}. \label{eq:rho}
\end{align}
$\rho_{0i}$ the initial density associated with $\rho_i$ (notice that values of $\rho_i$ are always positives as it is demonstrated in Appendix \ref{ap.c}).
On the other hand, the scalar factor satisfies the expression in Appendix \ref {ap.b2}, getting the dimensionless Ricci scalar (see Appendix \ref{ap.b})
\begin{equation}
\tilde{R}\approx \gamma_{\eta},\label{eq:gama}
\end{equation}
where $R=\tilde{R}H_0^2$, being $H_0$ the Hubble constant. 
This result indirectly calculates the Ricci scalar only constraining the value of $\gamma_n$. 

Also, when \eqref{ecdif} is solved (see details in Appendix \ref{ap.b}) we extract the solution \eqref{C}.
Thus, the dimensionless Friedmann equation takes the form of (see Appendix \ref{ap.c})
\begin{eqnarray}\label{eq:E2}
E(z)^2\beta^2= \Omega_{0m}\left[(1+z)^{3}-1\right]+\nonumber\\
\frac{3\Omega_{0r}}{4}\left[(1+z)^{4}-1\right]+\beta^2,\label{hf}
\end{eqnarray}
where $z$ emulates the redshift and $\beta$ satisfy \eqref{nu}.
Here, $c_1$ is a constant of integration and $\nu$ refers to the proximity of the Hubble parameter in the $f_{nc}$ cosmology with respect to the $\Lambda$CDM model at the actual time. The Friedmann constraint is already satisfied in Eq. \eqref{hf}. Finally, the 3/4 term in the Friedmann equation does not cause important effects in recent cosmology, as is shown in the Results section. However, a thorough analysis of the radiation epoch and the decrease in the contribution of radiation is necessary.

Moreover, the deceleration parameter reads as
\begin{eqnarray}
    q(z)=\frac{1}{2E(z)^2}\bigg[\frac{3\Omega_{0m}}{\beta^2}(1+z)^3+\frac{3\Omega_{0r}}{\beta^2}(1+z)^4\bigg]-1,
\end{eqnarray}
and the effective Equation of State (EoS) can be constructed with the formula $3w_{eff}(z)=2q(z)-1$.

Finally, the parameter space of the $f_{nc}(R)$ cosmology is defined as:
\begin{equation}
    \Theta = \{h, \Omega_{0b}, \Omega_{0m}, \beta\},
\end{equation}
where:
\begin{itemize}
    \item $h \equiv H_0/100$ is the dimensionless Hubble parameter,
    \item $\Omega_{0b}$ and $\Omega_{0m}$ represent the present-day density parameters for baryons and total matter (including baryons), respectively,
    \item $\beta$ is a free parameter of the modified gravity model.
\end{itemize}

%‰%%%‰%%%%%%%%%%%%%%%%%%%‰%%%‰%%%%%%%%%%%%%%%%%%
\section{Datasets and priors}
\label{Datasets}
%‰%%%‰%%%%%%%%%%%%%%%%%%%‰%%%‰%%%%%%%%%%%%%%%%%%

We constrain these cosmological parameters using the following observational datasets:
\begin{itemize}
    \item Cosmic Chronometers (CC). We use 33 Hubble parameter measurements $H(z)$ that span $0.07 < z < 1.965$ \cite{M_Moresco_2012, Moresco_2015, Moresco_2016, Moresco_2020, Jiao_2023, Tomasetti_2023}. The sample consists of 15 correlated measurements $H(z)$ and 18 uncorrelated data points, thus the $\chi^2$-function is built as
    \begin{equation}
        \chi^2_{\rm CC} = \sum_i^{18}\left( \frac{H_{obs}^i-H_{th}(z_i)}{\sigma^i}\right)^2 + \vec{H}\,{\rm Cov}^{-1}\vec{H}^{T},
    \end{equation}
    where the sum runs over all the uncorrelated sample, while the second term operates over the correlated sample, being Cov$^{-1}$ the inverse of the covariance matrix of the vector $\vec{H}$.
    \item Type Ia supernovae (SNIa). The Pantheon+ sample \cite{Scolnic2018-qf, Brout_2022} provides 1701 measurements of the correlated distance modulus across $0.001 < z < 2.26$. We analyze them using a correlated $\chi^2$ approach \cite{Conley2010} to marginalize nuisance parameters defined as
    \begin{equation}
        \chi^2_{\rm SNIa} = a + \log \left( \frac{e}{2\pi} \right) - \frac{b^2}{e},
    \end{equation}
    where 
    \begin{eqnarray}
        a &=& \Delta\boldsymbol{\tilde{\mu}}^{T}\cdot\mathbf{Cov_{P}^{-1}}\cdot\Delta\boldsymbol{\tilde{\mu}}, \nonumber\\
        b &=& \Delta\boldsymbol{\tilde{\mu}}^{T}\cdot\mathbf{Cov_{P}^{-1}}\cdot\Delta\mathbf{1}, \\
        e &=& \Delta\mathbf{1}^{T}\cdot\mathbf{Cov_{P}^{-1}}\cdot\Delta\mathbf{1}\, , \nonumber
    \end{eqnarray}
    and $\Delta\boldsymbol{\tilde{\mu}}$ is the vector of the difference between the theoretical and observed distance modulus defined as
        \begin{equation}
    \mu_{th}(z, \Theta) = 5 \log_{10} \left [ \frac{d_L(z)}{1\,{\rm Mpc}}\right] + 25,
    \end{equation}
    which is related to the luminosity distance $d_L$,  
        \begin{equation}\label{eq:dL}
    d_L(z)=(1+z)c\int_0^z\frac{dz^{\prime}}{H(z^{\prime})}\,,
    \end{equation}
    being $c$ the speed of light. Additionally, $\Delta\mathbf{1}=(1,1,\dots,1)^T$ is the transpose of the unit vector and $\mathbf{Cov_{P}}$ is the covariance matrix.
    
    \item Hydrogen II galaxies (HIIG). Our sample contains 181 distance modulus measurements of compact star-forming systems ($M < 10^9 M_{\odot}$) coming from HIIG and covering $0.01 < z < 2.6$ \cite{GonzalezMoran2019, Gonzalez-Moran:2021drc}. The $\chi^2$-function is built as
    \begin{equation}\label{eq:chi2_HIIG}
       \chi^2_{{\rm HIIG}}=\sum_i^{181}\frac{[\mu_{th}(z_i, {\Theta})-\mu_{obs}^i]^2}{\epsilon_i^2}.
    \end{equation}
    here $\mu_{obs}^i \pm \epsilon_i$ is the distance modulus and its uncertainty observed at the redshift $z_i$ and $\mu_{th}$ is its theoretical quantity.
    
    \item Intermediate-luminosity quasars (QSO). We analyze 120 angular size measurements from ultra-compact radio sources  in $0.462 < z < 2.73$ \cite{ShuoQSO:2017}. Due to their negligible dependence on redshift and intrinsic luminosity, they represent a fixed comoving-length ($l_m$) of standard ruler. The angular size $\theta$ is related to the angular diameter distance $D_A(z)$ and the intrinsic length of the QSO, $l_m$, through the relation $\theta(z)=l_m/D_A(z)$. For this work, we fix $l_m = 11.03\pm 0.25\,$pc \cite{ShuoQSO:2017}. An uncorrelated $\chi^2$ estimator is used as
    \begin{equation}
    \chi^2_{\rm QSO} = \sum_i^{120} \left( \frac{\theta_{obs}(z)-\theta_{th}}{\sigma_{\theta_{obs}}}\right)^2\,,
    \end{equation}
    where $\theta_{obs}(z) \pm \sigma_{\theta_{obs}}$ is the observed measurement  and it's corresponding uncertainty at redshift $z$, and $\theta_{th}$ is the theoretical counterpart.
    \item Baryon Acoustic Oscillations (BAO). The BAO signal provides measurements of the dilation scale ratio $D_V(z)/r_d$, where $D_V(z)$ is the volume-averaged distance at redshift $z$ and $r_d \equiv r_s(z_d)$ is the sound horizon at the drag epoch, the comoving angular diameter distance ratio $D_M(z)/r_d$, the Hubble distance ratio $D_H(z)/r_d \equiv (c/H(z))/r_d$, where $r_d$ is given by
    \begin{equation}
        r_d = \int_{z_d}^\infty \frac{c_s(z)dz}{H(z)}\,.
    \end{equation}  
    where $c_s(z)$ is the sound speed, and we use $z_d=1089.80\pm0.21$ \cite{Planck:2018}. The dilation scale is defined as \cite{Wigglez:Eisenstein2005}
    \begin{equation}
        D_V(z) = \sqrt[3]{z\,D_H(z)\,D_M^2(z)}\,.
    \end{equation}
    For a flat geometry, we have
    \begin{equation}
        D_M(z) = c\int_0^z\frac{dz'}{H(z')}\,.
    \end{equation}
    Thus, we consider a $\chi^2$-function as
    \begin{equation}
        \chi^2_{\rm BAO} = \sum_i^{12} \left( \frac{O_{th}(z_i)-O_{obs}^i}{\sigma^i}\right)^2,
    \end{equation}
    where $O_{obs}^i \pm \sigma^i$ is the measurement and its corresponding uncertainty at redshift $z^i$, and $O_{th}^i$ is its theoretical counterpart. In this work, 
    we analyze 12 BAO measurements from the first year of the Dark Energy Spectroscopic Instrument (DESI-DR1) data, spanning $0.1 < z < 4.16$ \cite{desicollaboration2024desi}.
    
    \item Fast radio bursts (FRBs). FRBs are extremely bright, millisecond-duration radio transients that serve as valuable probes for cosmological studies \cite{ZhouFRB:2014}. We utilize a new sample of 92 localized FRBs from \cite{wang2025:FRB}, spanning a redshift range of $0.0385 < z < 1.354$. To contrast model with data, we built a $\chi^2$-function as
    \begin{equation}
        \chi^2_{\rm FRB} = \sum_i^{92}\left( \frac{{\rm DM}_{\rm IGM}^{th}(z_i) - {\rm DM}_{\rm IGM}^{obs}}{\sigma^i} \right)^2,
    \end{equation}
    having ${\rm DM} un_{\rm IGM}^{obs} \pm \sigma^i$ is the dispersion measurement and its uncertainty at redshift $z_i$ of the FRB signal when travels through the intergalactic ionized medium. ${\rm DM}_{\rm IGM}^{obs}$ is obtained as
    \begin{equation}
        {\rm DM}_{\rm IGM}^{obs} = {\rm DM}_{\rm total} - {\rm DM}_{\rm MW} - {\rm DM}_{\rm halo} - \frac{{\rm DM}_{\rm host}}{1+z},
    \end{equation}
    where ${\rm DM}_{\rm total}$ is the total observation of the medium dispersion, and ${\rm DM}_{\rm i}$ is the contribution from the Milky Way (MW), the galactic halo (halo), and the host galaxies (host). 
    The theoretical counterpart is estimated as
    \begin{equation}
        {\rm DM}_{\rm IGM}^{th} = \frac{3c\Omega_{0b}H^2_0f_{\rm IGM}}{8\pi G m_p}\int_0^z \frac{\chi(z')(1+z')}{H(z')}dz',
    \end{equation}
    where $\Omega_{0b}$ is the baryon density parameter, $m_p$ is the proton mass, $f_{\rm IGM}\approx 0.84$ is the fraction of baryons in the IGM, and $\chi(z)=7/8$ is the ionization fractions of intergalactic ionized hydrogen and helium. For more details see \cite{wang2025:FRB}.
\end{itemize}

Our analysis employs a Markov Chain Monte Carlo (MCMC) approach implemented through the \texttt{emcee} Python package \citep{Foreman:2013}. To ensure convergence, we
monitor the autocorrelation function at each 10 steps of the process. After validating fluctuations of the autocorrelation function of the order of $10^{-3}$, we get 4000 chains of 200 steps each to sample the space $\Theta$, with Gaussian priors in $h = 0.6766 \pm 0.0042$, $\Omega_{0b}h^2 = 0.02202 \pm 0.00046$, $\Omega_{0m} = 0.3111 \pm 0.0056$ \cite{Planck:2018} and the uniform prior in $\beta \in (0, 3)$. It is worth to mention that due to the high degeneracy between $\beta$ and $h$ and $\Omega_{0m}$, it is required to establish Gaussian priors on $h$ and $\Omega_{0m}$, allowing $\beta$ to vary freely on the mentioned range.

%%%%%%%%%%%%%%%%%%%%%%%%%%%%%%%%%%%%%%%%%%%%
\section{Results}\label{Results}
%%%%%%%%%%%%%%%%%%%%%%%%%%%%%%%%%%%%%%%%%%%%

\begin{figure*}
    \centering
    \includegraphics[width=0.8\textwidth]{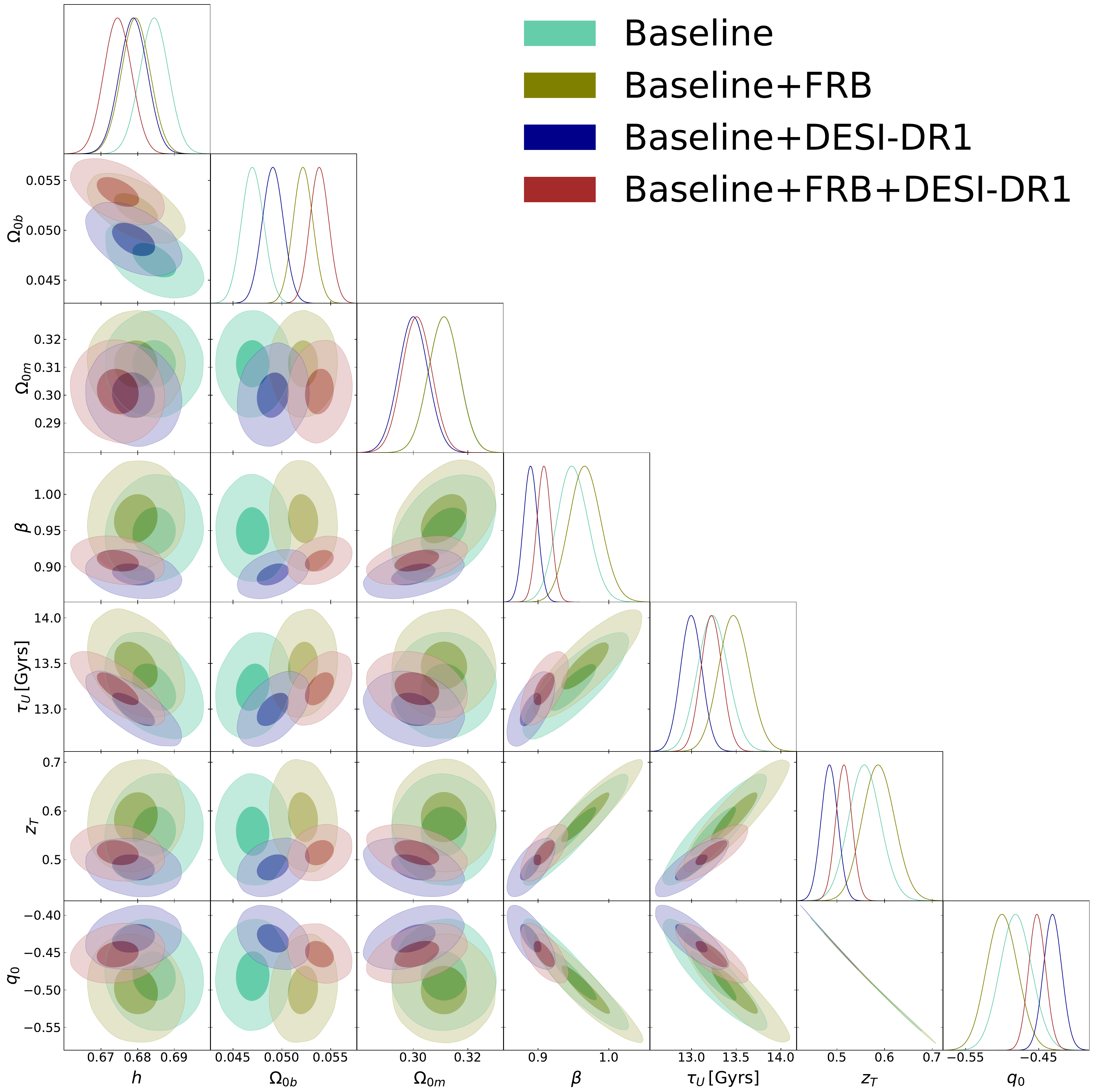
    }
    \caption{1D posterior distributions and 2D contours at $1\sigma$ (inner region) and $3\sigma$ (outermost region) CL for $f_{nc}(R)$ model.}
    \label{fig:contours}
\end{figure*}

The parameter space $\Theta$ of the model is constrained using a combination of the cosmological samples described in \ref{Datasets}. We construct a baseline sample comprising CC, SNIa, HIIG and QSO data, which together cover the late-time evolution of the Universe. To this baseline, we incorporate the BAO measurements from DESI-DR1. Additionally, we explore an extended sample by further including FRBs.
The results of the Bayesian analysis are presented in Table \ref{tab:bf_model}, which reports the median values of the model parameters along with the uncertainties $1\sigma$ (68\% confidence level, CL). From these constraints, we derive the age of the Universe ($\tau_U$), the redshift of the acceleration-deceleration transition ($z_T$), and the present-day deceleration parameter ($q_0$).
Furthermore, Figure \ref{fig:contours} shows the 2D confidence regions in $1\sigma$ (inner contour) and $3\sigma$ (99\% CL, outer contour), as well as the 1D posterior distributions for the parameters in $\Theta$.
First, due to the dependence of the baryon components in the FRB sample, the median values of $\Omega_{0b}$ are approximately $5\sigma$ higher than those obtained when this sample is excluded. Using Eq.~\eqref{nu} with the plus sign, we estimate $\nu = -0.100^{+0.043}_{-0.040}$, $-0.066^{+0.044}_{-0.042}$ for baseline and baseline+FRB, respectively. When DESI-DR1 is included, these values become: 
$\nu=-0.209^{+0.018}_{-0.017}$ and $-0.175^{+0.018}_{-0.017}$, respectively.  From Eq. \eqref{eq:gama}, we estimate the Ricci scalar as $\tilde{R} = 3.133^{+0.053}_{-0.051}$ and $\tilde{R} = 3.175^{+0.052}_{-0.051}$ for the baseline and baseline+FRB cases, respectively. These values change to $\tilde{R} = 2.988 \pm 0.025$ and $\tilde{R} = 3.035 \pm 0.024$ when BAO measurements are included. 

The contour plots show that the parameters $\beta$ and the actual time deceleration are anti-correlated, thus, when $\beta$ increases (decreases) the age of the Universe and $z_T$ changes accordingly. Furthermore, $\tau_U$ is correlated with $z_T$ and inversely related to $q_0$.
Furthermore, we find that the estimated age of the Universe is consistent with $\Lambda$CDM predictions within $2\sigma$ for samples excluding FRB data. This agreement deteriorates significantly when FRB data are included, with the discrepancy increasing up to $4.1\sigma$. A similar trend is observed for both parameters, $z_T$ and $q_0$, suggesting that the FRB data influence these cosmological parameters. 
Moreover, we identify correlations between the characteristic parameter $\beta$ of the $f_{nc}(R)$ model and these observables: $\beta$ exhibits a positive correlation with $\tau_U$ and $z_T$, but a negative correlation with $q_0$.

\begin{table*}[ht!]
	\centering
	\caption{Median values and their $1\sigma$ confidence interval for the $f_{nc}(R)$ cosmology and $\Lambda$CDM.}
	\label{tab:bf_model}
	\begin{tabular}{lccccccccc} 
    \hline
    Data & $\chi^2$  & $h$ & $\Omega_{0b}$ & $\Omega_{0m}$ & $\beta$ & $\tau_U \,[\rm{Gyrs}]$  & $z_T $  &  $q_0 $  \\ [0.9ex] 
    \hline
    \multicolumn{9}{c}{$f_{nc}(R)$ cosmology} \\ [0.9ex]
Baseline &5632.20 & $0.684^{+0.004}_{-0.004}$  & $0.047^{+0.001}_{-0.001}$  & $0.311^{+0.006}_{-0.006}$  & $0.949^{+0.022}_{-0.021}$  & $13.238^{+0.172}_{-0.168}$  & $0.558^{+0.034}_{-0.033}$  & $-0.481^{+0.022}_{-0.022}$  \\ [0.9ex] 
Baseline+FRB &6063.62 & $0.679^{+0.004}_{-0.004}$  & $0.052^{+0.001}_{-0.001}$  & $0.311^{+0.006}_{-0.006}$  & $0.966^{+0.023}_{-0.022}$  & $13.475^{+0.175}_{-0.167}$  & $0.587^{+0.034}_{-0.033}$  & $-0.500^{+0.021}_{-0.021}$  \\ [0.9ex] 
Baseline+DESI-DR1 &5688.76 & $0.679^{+0.004}_{-0.004}$  & $0.049^{+0.001}_{-0.001}$  & $0.300^{+0.005}_{-0.005}$  & $0.889^{+0.010}_{-0.010}$  & $12.996^{+0.120}_{-0.118}$  & $0.484^{+0.018}_{-0.017}$  & $-0.431^{+0.013}_{-0.012}$  \\ [0.9ex] 
Baseline+FRB+DESI-DR1 &6110.20 & $0.675^{+0.004}_{-0.004}$  & $0.054^{+0.001}_{-0.001}$  & $0.301^{+0.005}_{-0.005}$  & $0.908^{+0.010}_{-0.010}$  & $13.223^{+0.118}_{-0.117}$  & $0.515^{+0.017}_{-0.017}$  & $-0.452^{+0.012}_{-0.011}$  \\ [0.9ex]
% \hline
\hline
        \multicolumn{9}{c}{$\Lambda$CDM cosmology} \\ [0.9ex]
Baseline &5643.60 & $0.685^{+0.004}_{-0.004}$  & $0.047^{+0.001}_{-0.001}$  & $0.316^{+0.006}_{-0.006}$ & - & $13.575^{+0.111}_{-0.109}$  & $0.630^{+0.015}_{-0.015}$  & $-0.526^{+0.009}_{-0.009}$  \\ [0.9ex] 
Baseline+FRB &6071.90 & $0.680^{+0.004}_{-0.004}$  & $0.052^{+0.001}_{-0.001}$  & $0.314^{+0.006}_{-0.006}$ & -  & $13.691^{+0.110}_{-0.110}$  & $0.634^{+0.015}_{-0.015}$  & $-0.528^{+0.009}_{-0.009}$  \\ [0.9ex] 
Baseline+DESI-DR1 &5668.15 & $0.681^{+0.004}_{-0.004}$  & $0.049^{+0.001}_{-0.001}$  & $0.316^{+0.006}_{-0.006}$ & -  & $13.655^{+0.107}_{-0.105}$  & $0.630^{+0.014}_{-0.014}$  & $-0.526^{+0.009}_{-0.008}$  \\ [0.9ex] 
Baseline+FRB+DESI-DR1 &6088.90 & $0.679^{+0.004}_{-0.004}$  & $0.052^{+0.001}_{-0.001}$  & $0.313^{+0.006}_{-0.006}$ & -  & $13.723^{+0.107}_{-0.106}$  & $0.636^{+0.014}_{-0.014}$  & $-0.530^{+0.008}_{-0.008}$  \\ [0.9ex]
\hline
\hline
        \multicolumn{9}{c}{$f_{nc}(R)$ cosmology (Uniform priors)}\\ [0.9ex]
Baseline &5603.51 & $0.742^{+0.012}_{-0.012}$  & $0.092^{+0.061}_{-0.061}$  & $0.623^{+0.266}_{-0.338}$  & $1.379^{+0.266}_{-0.445}$  & $12.402^{+0.219}_{-0.211}$  & $0.600^{+0.035}_{-0.034}$  & $-0.508^{+0.021}_{-0.021}$  \\ [0.9ex] 
Baseline+FRB &9486.19 & $0.720^{+0.012}_{-0.011}$  & $0.063^{+0.001}_{-0.001}$  & $0.625^{+0.264}_{-0.336}$  & $1.203^{+0.232}_{-0.385}$  & $11.800^{+0.195}_{-0.190}$  & $0.380^{+0.029}_{-0.029}$  & $-0.352^{+0.024}_{-0.023}$  \\ [0.9ex] 
Baseline+DESI-DR1 &5620.99 & $0.744^{+0.012}_{-0.011}$  & $0.140^{+0.030}_{-0.054}$  & $0.503^{+0.135}_{-0.193}$  & $1.265^{+0.154}_{-0.272}$  & $12.502^{+0.210}_{-0.206}$  & $0.630^{+0.027}_{-0.027}$  & $-0.526^{+0.017}_{-0.016}$  \\ [0.9ex] 
Baseline+FRB+DESI-DR1 &9534.09 & $0.730^{+0.012}_{-0.011}$  & $0.061^{+0.001}_{-0.001}$  & $0.311^{+0.019}_{-0.018}$  & $0.899^{+0.018}_{-0.017}$  & $12.035^{+0.201}_{-0.195}$  & $0.472^{+0.025}_{-0.025}$  & $-0.422^{+0.018}_{-0.018}$  \\ [0.9ex] 
\hline
\hline
	\end{tabular}
\end{table*}

\begin{figure*}
    \centering
    \includegraphics[width=0.31\textwidth]{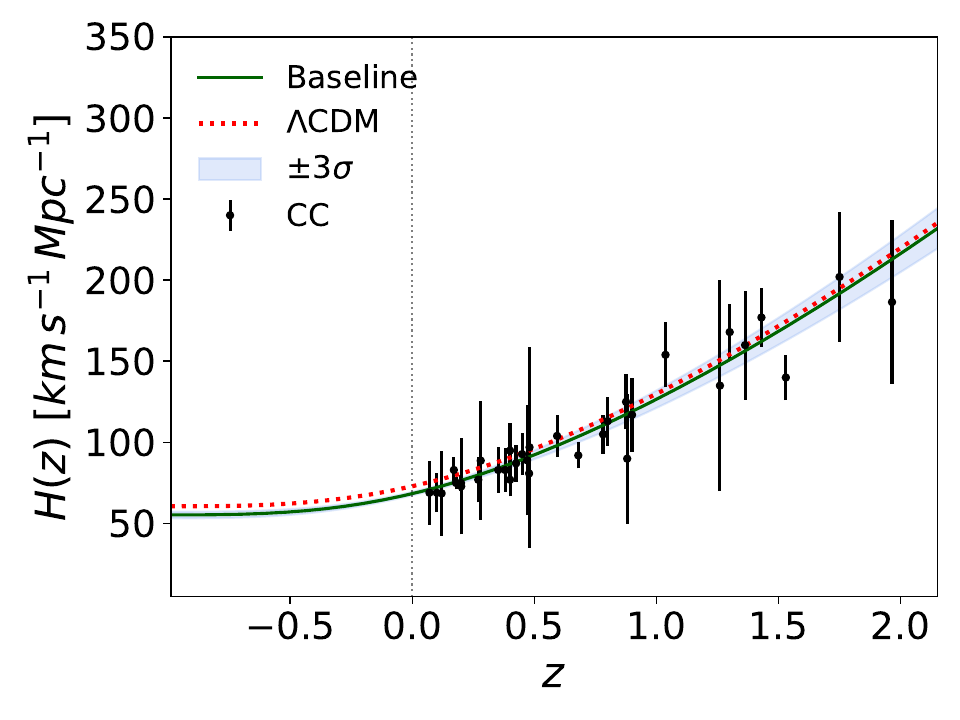}
    \includegraphics[width=0.31\textwidth]{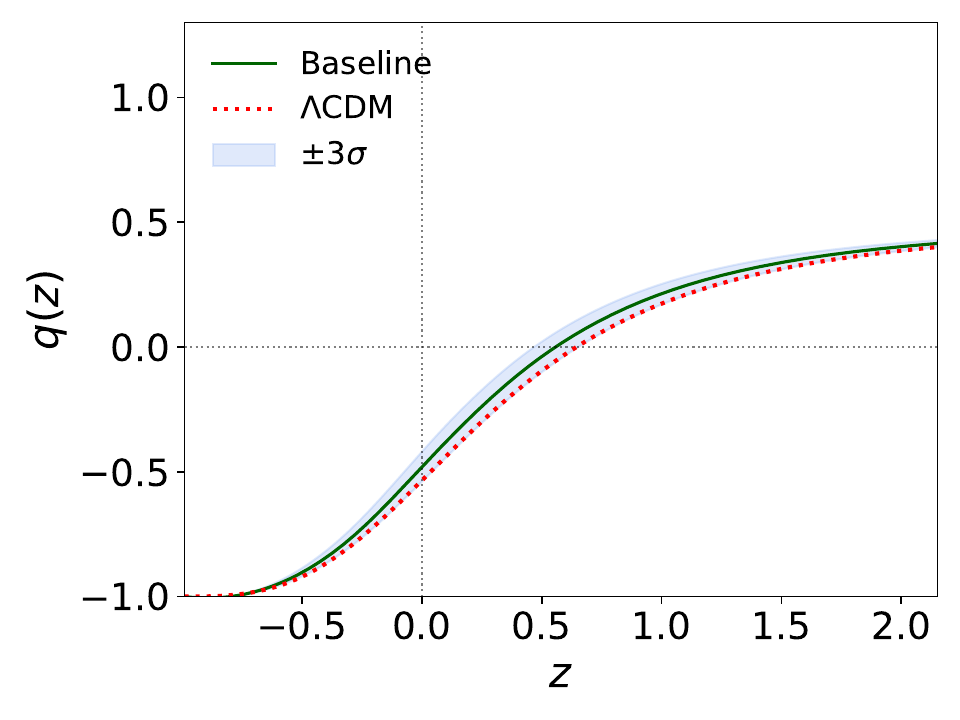}
    \includegraphics[width=0.31\textwidth]{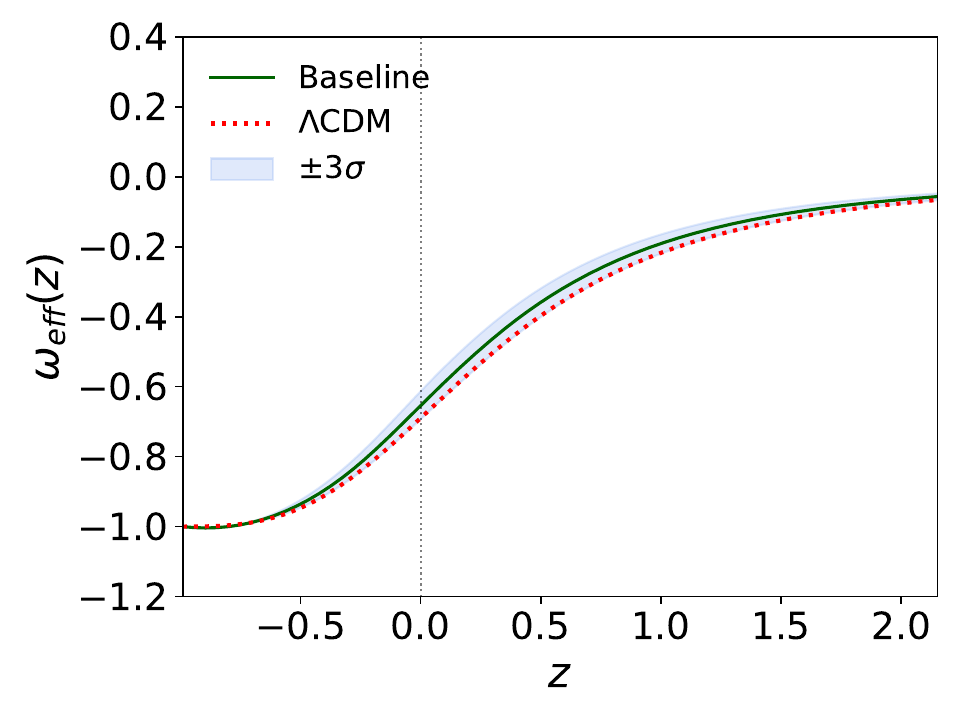}\\
    \includegraphics[width=0.31\textwidth]{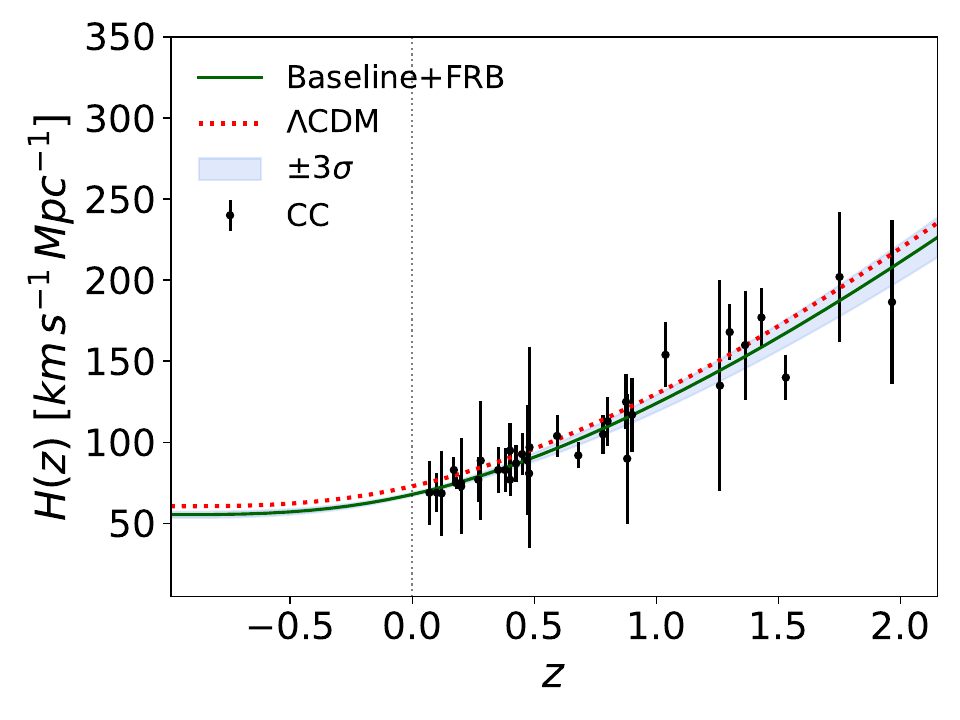}
    \includegraphics[width=0.31\textwidth]{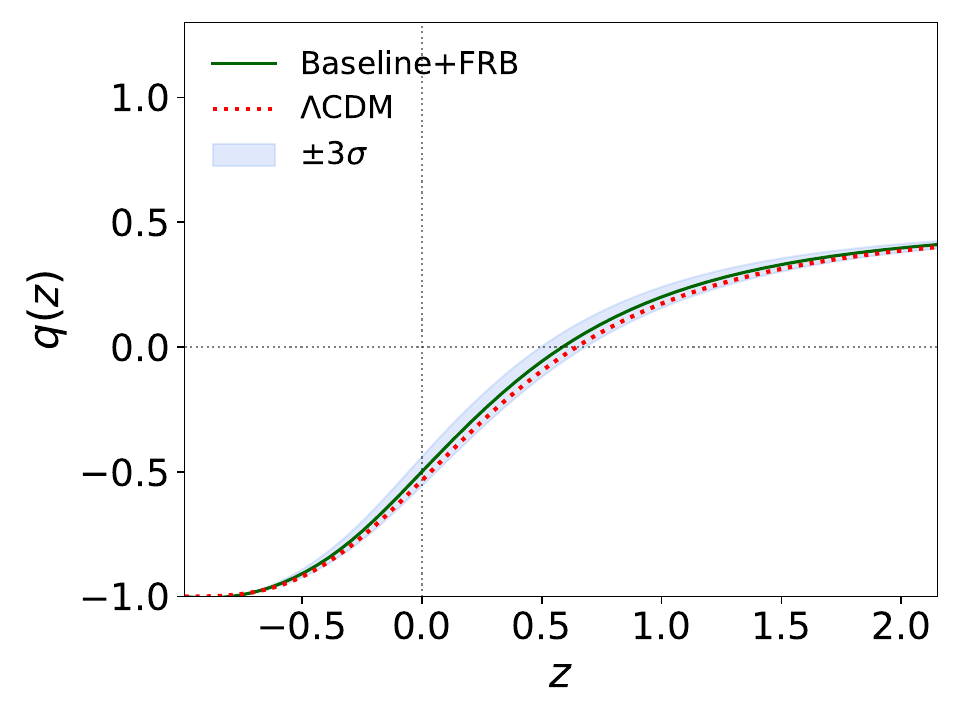}
    \includegraphics[width=0.31\textwidth]{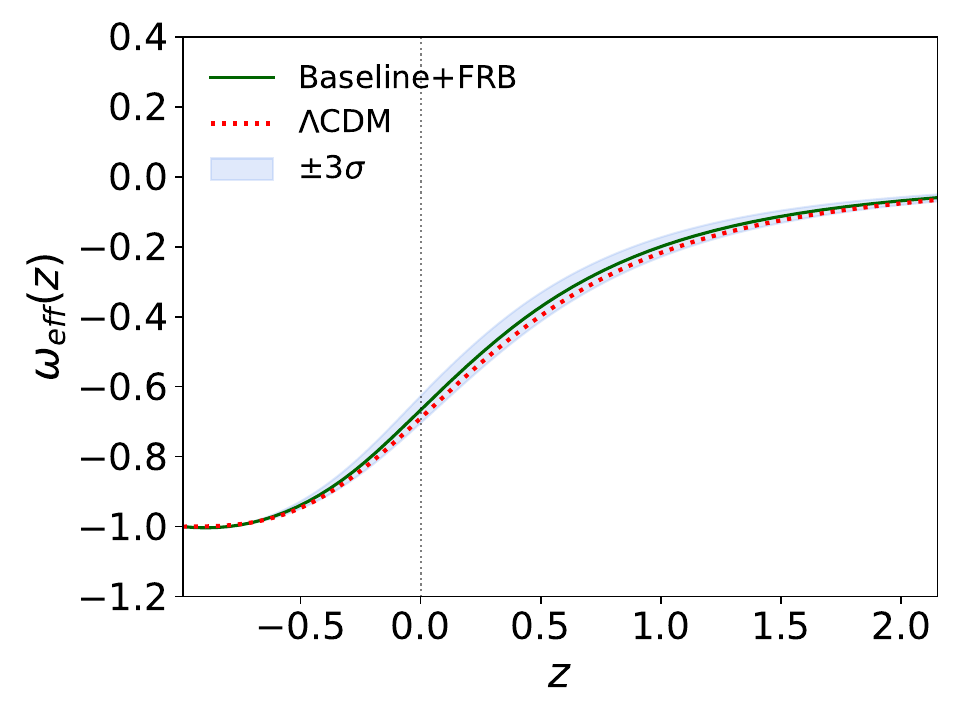}\\
    \includegraphics[width=0.31\textwidth]{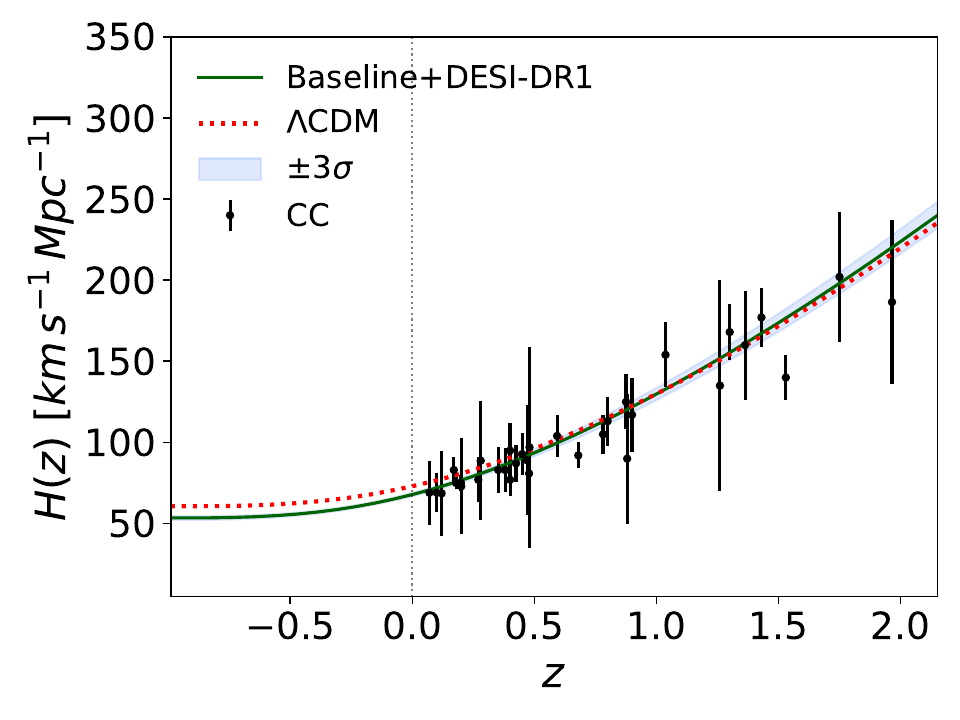}
    \includegraphics[width=0.31\textwidth]{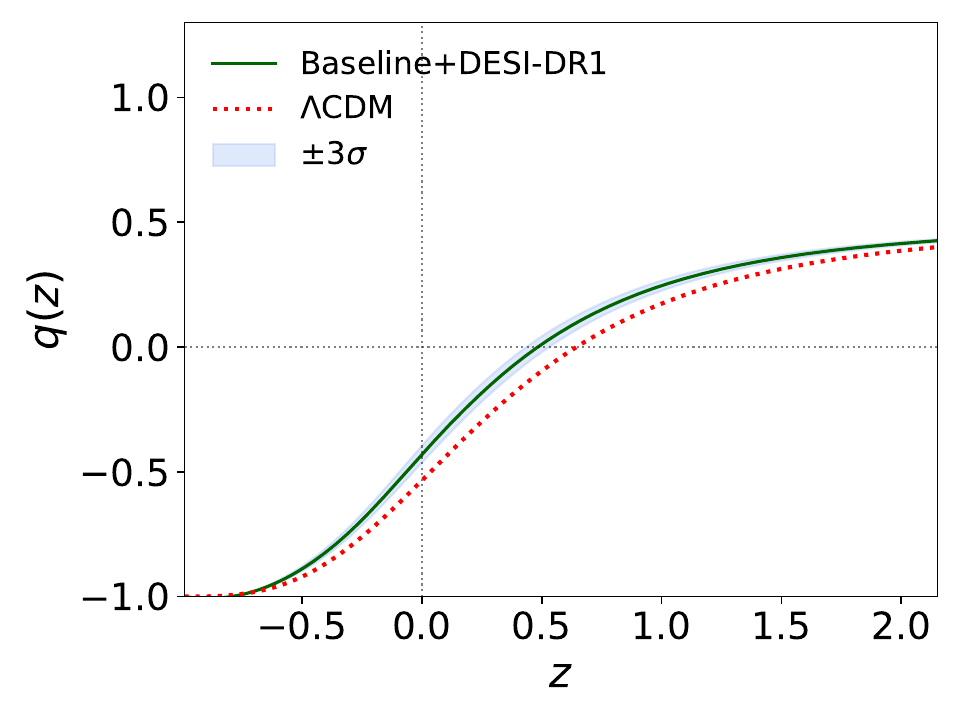}
    \includegraphics[width=0.31\textwidth]{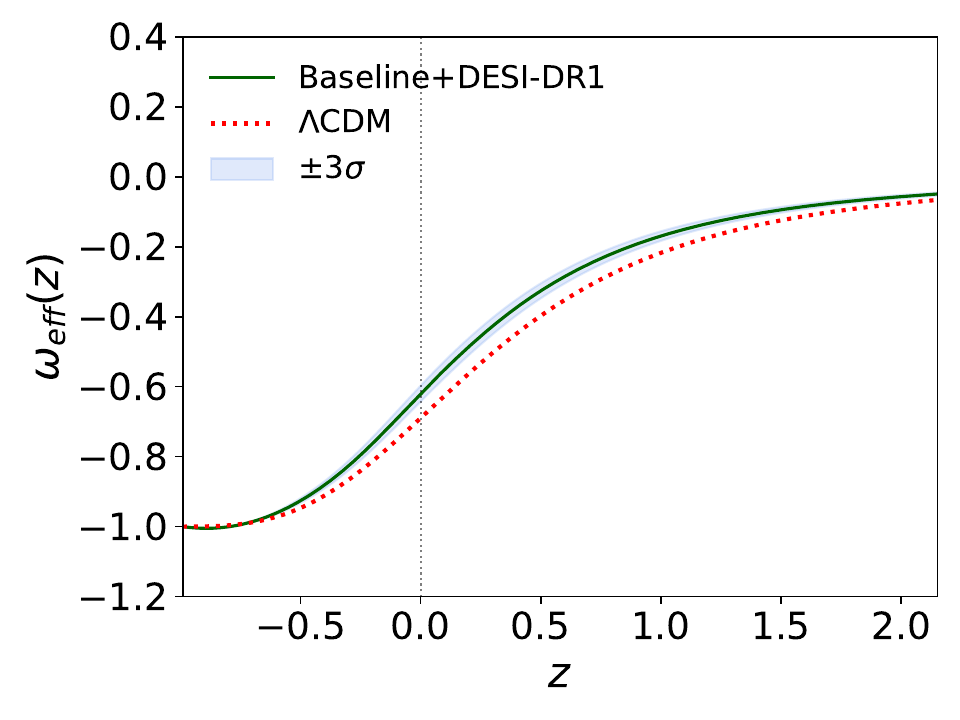}\\
    \includegraphics[width=0.31\textwidth]{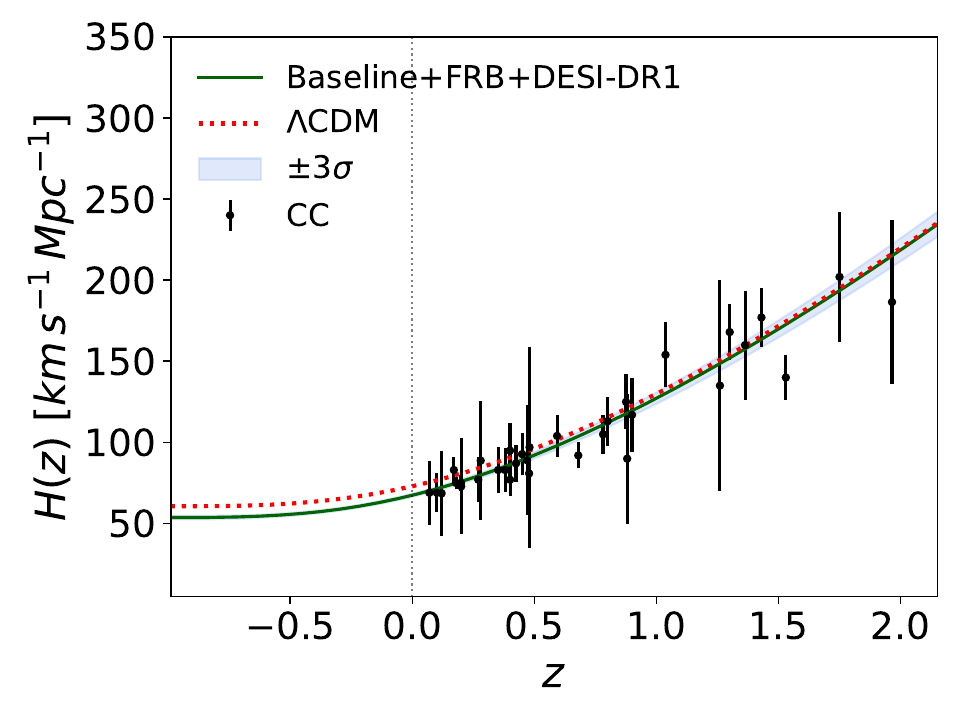}
    \includegraphics[width=0.31\textwidth]{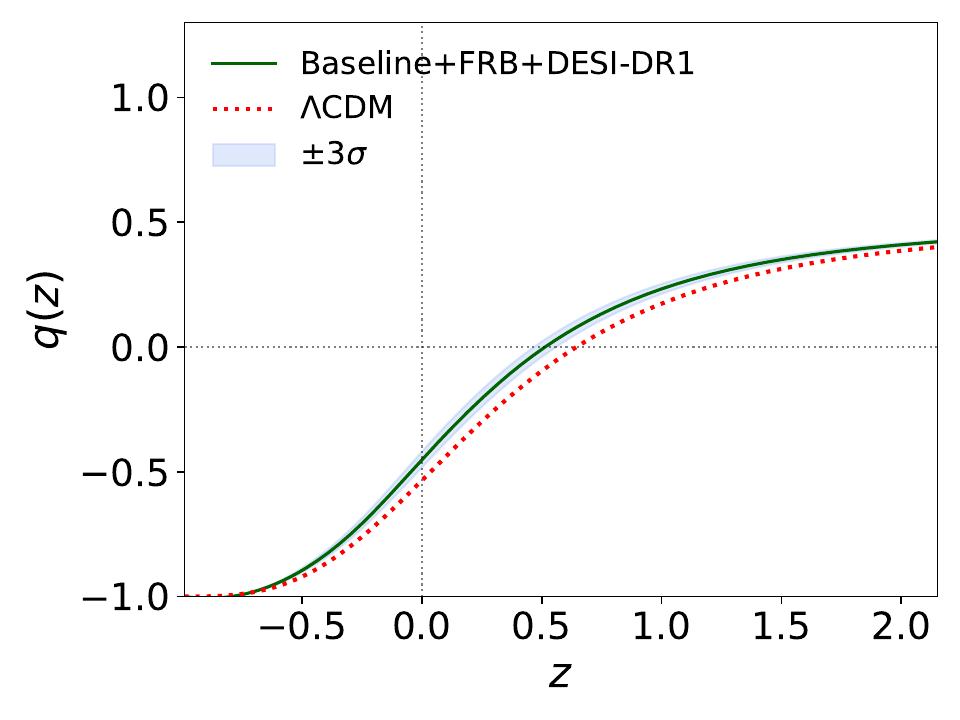}
    \includegraphics[width=0.31\textwidth]{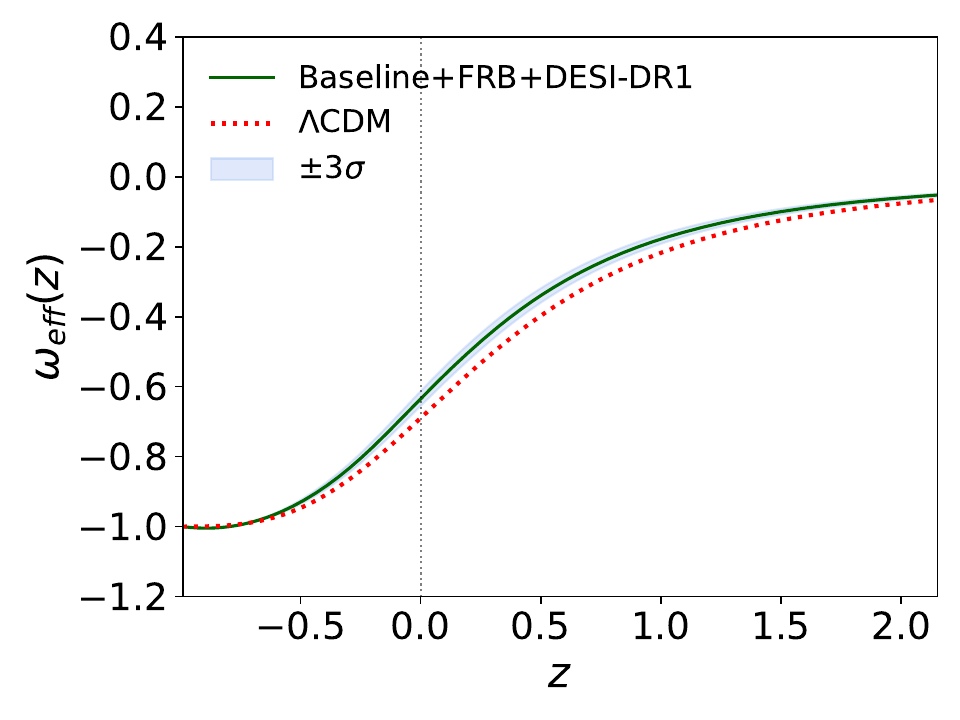}\\
    \caption{Reconstruction of the Hubble parameter (left column), the deceleration parameter (middle column) and the effective EoS of the Universe for the $f_{nc}(R)$ cosmology in the redshift range $-1<z<2.2$ using different data combinations. The standard $\Lambda$CDM model is included as red dashed lines.}
    \label{fig:cosmography}
\end{figure*}

Figure~\ref{fig:cosmography} illustrates the reconstructed evolution of the Hubble parameter $H(z)$, the deceleration parameter $q(z)$ and the effective equation of state $w_{\rm eff}(z)$, along with their respective confidence regions $3\sigma$, which span the redshift range $-1 < z < 2.2$ for various combinations of data. Although the general trends of these cosmological functions align with the predictions of the $\Lambda$CDM model when the DESI-DR1 data are excluded, notable deviations (exceeding $3\sigma$) are observed when the DESI-DR1 sample is incorporated into the analysis.

Furthermore, we perform a statistical model selection using two of the widely used metrics to compare both cosmological models, the $f_{nc}(R)$ model and the $\Lambda$CDM, based on their goodness of fit and complexity. The Akaike Information Criterion (AIC) aims to avoid overfitting by penalizing models with more parameters and is defined as \cite{AIC:1974, Sugiura:1978} ${\rm AIC}=2k - 2 \log (\hat{L})$, where $k$ is the number of parameters in the model and $\hat{L}$ is the maximum value of the likelihood function for the model, which is related to the chi-square function $\chi^2$ via $- 2 \log (\hat{L}) = \chi^2$. Lower AIC values indicate a better trade-off between model fit and simplicity. AIC is interpreted through the difference between two models, the preferred model being the one with the lowest AIC value, with the following rules. The values of $\Delta$AIC$< 4$ mean that both models are equally supported by the data. If the difference is $4 < \Delta$AIC$< 10$, the data still support the given model but less strongly than the preferred one. Finally, if $\Delta$AIC$> 10$, the data do not support the given model.

The second metric, named the Bayesian Information Criterion (BIC), is defined as \cite{schwarz1978} ${\rm BIC} =k\log(N_s) - 2 \log (\hat{L})$ where $N_s$ is the number of data points.  BIC imposes a heavier penalty on models with more parameters than AIC, especially as the sample size $N_s$ increases. Similarly to AIC, lower BIC values indicate a better model. The interpretation of the BIC difference ($\Delta$BIC) is as follows. If $\Delta$BIC $< 2$, there is no significant evidence against the model. If $2 < \Delta$BIC $< 6$, there is modest evidence against the candidate model. If $6 < \Delta$BIC $< 10$, the evidence against the model is strong. Finally, if $\Delta$BIC $> 10$, the evidence against the model is very strong.

Based on these criteria, we conclude the following. We do not find evidence against the $f_{nc}(R)$ cosmology in the late-time phase of the universe (considering both the baseline and the baseline+FRB samples). However, the strongest evidence against this model emerges when the BAO measurements are included.

\begin{table*}[ht!]
\centering
\caption{Statistical comparison between  $f_{nc}(R)$ model and $\Lambda$CDM for several datasets using AIC and BIC. Negative values of $\Delta$ favor $f_{nc}(R)$ cosmology.}
\begin{tabular}{lcccccc}
\hline
DATA & AIC($f_{nc}(R)$) & AIC(LCDM) & $\Delta$AIC & BIC($f_{nc}(R)$) & BIC(LCDM) & $\Delta$BIC \\
\hline
Baseline & 5640.20 & 5649.60 & -9.40 & 5662.67 & 5666.45 & -3.78 \\
Baseline + DESI-DR1 & 5696.76 & 5674.14 & 22.62 & 5719.25 & 5691.01 & 28.24 \\
Baseline + FRB & 6071.62 & 6077.90 & -6.28 & 6094.26 & 6094.88 & -0.62 \\
Baseline + FRB + DESI-DR1 & 6118.20 & 6094.90 & 23.30 & 6140.87 & 6111.90 & 28.97 \\
\hline
\end{tabular}
\label{tab:aic_bic}
\end{table*}

\begin{figure*}
    \centering
    \includegraphics[width=0.8\textwidth]{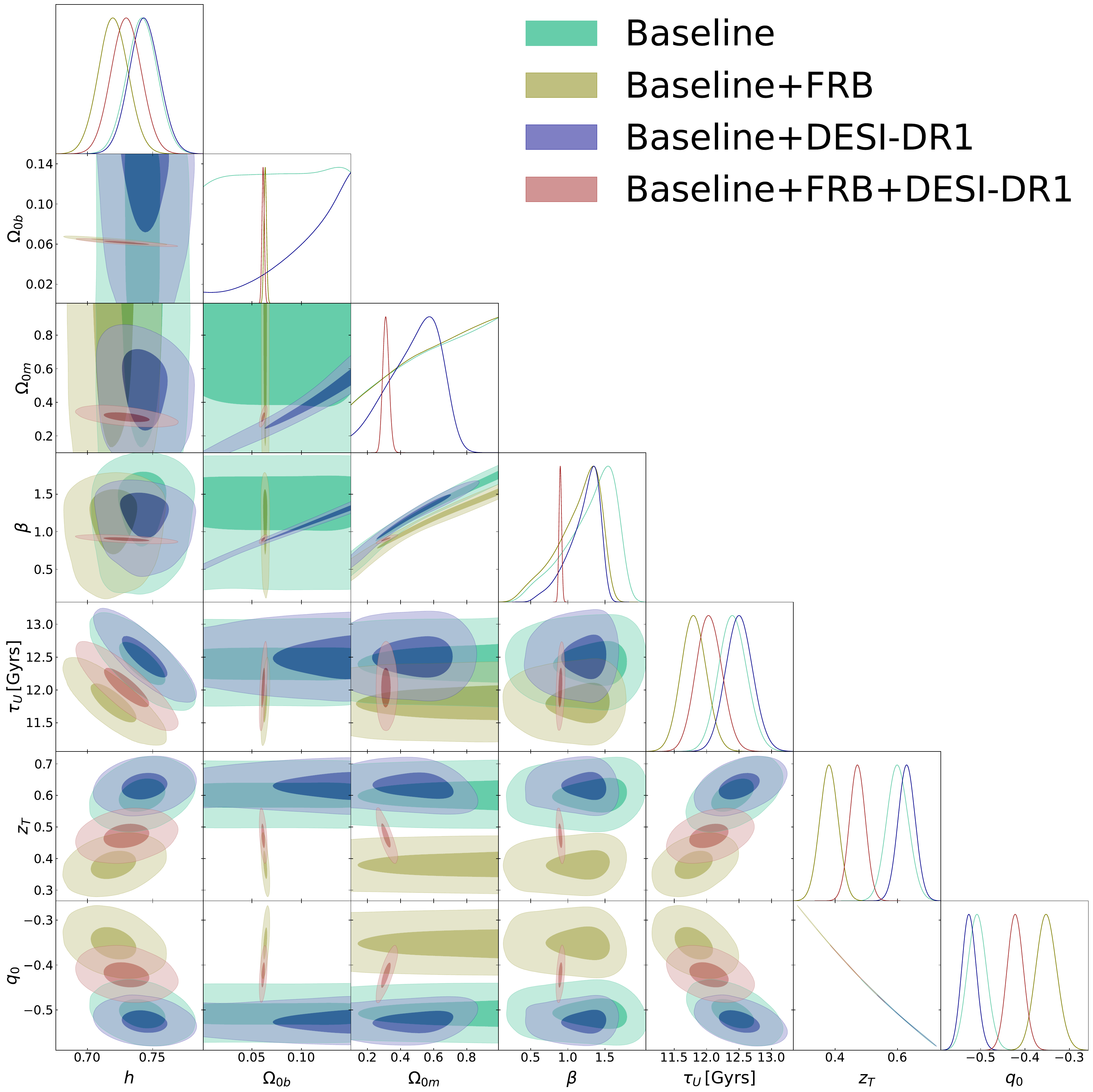}
    \caption{1D posterior distributions and 2D contours at $1\sigma$ (inner region) and $3\sigma$ (outermost region) CL for $f_{nc}(R)$ model considering Uniform priors on $0.2<h<1$ and $0.1<\Omega_{0m}<1$.}
    \label{fig:contours_flat}
\end{figure*}

Finally, Fig. \ref{fig:contours_flat} displays the 2D contours and the 1D posterior distributions when uniform priors are considered on $0.1<h<1$ and $0.1<\Omega_{0m}<1$. Furthermore, Table \ref{tab:bf_model} presents the median and their uncertainties at $1\sigma$. The panel $\Omega_{0m}-\beta$ shows the strong correlation between both parameters, as presented in equation \eqref{eq:E2}. As a possible consequence, we find that the age of the Universe is younger than the age of the oldest globular cluster ($\tau_{GC}=13.39 \pm 0.1\,({\rm stat.}) \pm 0.23 \, ({\rm sys.})\,$Gyrs. ) \cite{valcin2025}, finding deviations up to $4.2\sigma$. This deviation is reduced to $1.4\sigma$ when Gaussian priors on Planck values are considered. With respect to the parameters $z_T$ and $q_0$, we find values in agreement with those reported with our nominal analysis (Gaussian priors) with deviations up to $2\sigma$.

%%%%%%%%%%%%%%%%%%%%%%%%%%%%%%%%%%%%%%%%%%%%
\section{Conclusions and Outlooks} \label{Conclusions}
%%%%%%%%%%%%%%%%%%%%%%%%%%%%%%%%%%%%%%%%%%%%%

This paper uses a $f_{nc}(R)$ gravity framework, where $f_{nc}$ is a Starobinsky-type function combined with coefficients that exhibit a non-commutative background.
We refer to Appendix \ref{ap.b2}
for the explicit form of $f_{nc}$, briefly discussing the main characteristics and the theoretical aspects behind its proposal. The $f_{nc}(R)$ cosmology is closely related to the mathematical formalism in Appendix \ref{ap.b}, which leads to the equation \eqref{hf} that allows us to study the dynamics in the model. Finally, the model is tested by implementing an MCMC likelihood analysis.

Initially, $\beta$ represents the solution to the differential equation \eqref{ecdif}, and the parameter $\nu$ is introduced to alleviate the actual time ratio between the Hubble parameter for the $f_{nc}$ Cosmology and the respective one in the $\Lambda$CDM model. That is, we ask the question. 
\begin{equation}
\beta H(z)\underset{z\to 0}{\sim}H_{\Lambda}
\qquad\text{or}\qquad \frac{\beta H(z)}{H_{\Lambda}(z)}=1\pm \nu,\label{ratio}
\end{equation}
stating that if $\vert z\vert<1$, then $\nu$ is small enough and $\beta$ is close to 1. Figure \ref{fig:contours} shows the strong relation between the Age of the Universe and $\beta$, taking into account the contribution of non-commutativity represented by this parameter and hence the function $f_{nc}$.

The set of parameters that emerge from the model is constrained by a baseline composed of CC, SNIa, HIIG and QSO and recent DESI and FRBs data samples. Using our constraints, we derive the age of the Universe, the redshift of the deceleration-acceleration transition, and the present deceleration parameter, all compiled in table \ref{tab:bf_model},  and find good agreement with the standard cosmological model. 
Additionally, the FRB sample shows a $3.5\sigma$ difference in $\Omega_{0b}$ compared to when it is not included in the sample. The main parameters of the theory are constrained, with $\nu = -0.100^{+0.043}_{-0.040}$, $-0.066^{+0.044}_{-0.042}$ for the baseline and baseline + FRB, respectively. When DESI-DR1 is included, the values of $\nu$ are $\nu=-0.209^{+0.018}_{-0.017}$ and $-0.175^{+0.018}_{-0.017}$, respectively. 
Curiously, knowledge about the geometric properties of our Universe in this scenario can be extracted from Eq. \eqref{eq:gama}, which relates the factor $\gamma_\eta$ to the dimensionless Ricci scalar, with a value approximately equal to $\sim3$ or $R\approx3H_0^2$, compared to the standard model where $R^{\Lambda CDM}\approx 3\Omega_{0\Lambda}H_0^2$.
Finally, based on the AIC and BIC analyses, we conclude that there is no evidence against $f_{nc}(R)$ cosmology in late-time evolution, when considering the baseline+FRB data sample. However, when the BAO data sample is included, it provides the strongest evidence against the $f_{nc}(R)$ model. We believe that this strongest evidence against the model, when the BAO data sample is added, is due to the model dependence on $\Lambda$CDM. Another interesting point is related to the 3/4 term in the Friedmann equation associated with radiation. Although there are no important changes in the evolution of $H$ in late times, at early times the factor could cause interesting changes in the evolution of radiation, for example, in nucleosynthesis where finger prints of the model could be observed.

Additionally, Fig. \ref{fig:contours_flat} displays the 2D contours and the 1D posterior distributions with uniform priors. Table \ref{tab:bf_model} presents the median and their uncertainties at $1\sigma$. As we discussed in the section of results, we find that the age of the Universe is younger than the age of the oldest globular cluster ($\tau_{GC}=13.39 \pm 0.1\,({\rm stat.}) \pm 0.23 \, ({\rm sys.})\,$Gyrs. ) \cite{valcin2025}, finding deviations up to $4.2\sigma$. A reduction to $1.4\sigma$ is obtained when Gaussian priors on Planck values are applied. Regarding the parameters $z_T$ and $q_0$, we find values in agreement with those reported with our nominal analysis (Gaussian priors) with deviations up to $2\sigma$.

In this direction, we remark the ongoing interest in investigating with more detail whether the non-commutativity could be relevant in the late-time or early-time Universe, as is addressed in other models it has been shown that the role of non-
commutativity may have very important implications in Cosmology \cite{ncc.cos,ncc.cos1}, even when considering modified theories of gravity \cite{ncc.cos,ncc.wh,ncc.geo}. In particular, \cite{cyclic} present some implications of incorporating a non-commutative deformation in a $f(R)$ scenario, where the first solutions for the scale factor have a structure similar to previous solutions reported in the literature. Also, in other cases, the solutions obtained show that its
evolution has the behavior of a nonsingular cyclic universe, in
contrast to the respective commutative counterpart. However, this is beyond the scope of this paper and will be discussed in future work.

\begin{acknowledgments}
We thank anonymous referees for thoughtful remarks and suggestions. J.A.A.-M. acknowledges SECIHTI for the support by a postdoctoral fellowship at CINVESTAV, M\'exico.
A.H.A. thanks to the support from Luis Aguilar, 
Alejandro de Le\'on, Carlos Flores, and Jair Garc\'ia of the Laboratorio 
Nacional de Visualizaci\'on Cient\'ifica Avanzada. M.A.G.-A. acknowledges support from c\'atedra Marcos Moshinsky, Universidad Iberoamericana for the support with the National Research System (SNI) grant and the project 0056 from Universidad Iberoamericana: Nuestro Universo en Aceleraci\'on, energ\'ia oscura o modificaciones a la relatividad general. The numerical analysis was also carried out by {\it Numerical Integration for Cosmological Theory and Experiments in High-energy Astrophysics} (Nicte Ha) cluster at IBERO University, acquired through c\'atedra MM support. A.H.A, V.M. and M.A.G.-A acknowledge partial support from project ANID Vinculaci\'on Internacional FOVI240098. V.M. acknowledges support from Centro de Astrof\'{\i}sica de Valpara\'{\i}so CIDI 21.
\end{acknowledgments}

\bibliography{main-FLRW}

%%%%%%%%%%%%%%%%%%%%%%%%%%%%%%%%%%
\appendix 
%%%%%%%%%%%%%%%%%%%%%%%%%%%%%%%%%%

\section{Complementary notions for the $f_{nc}(R)$ Cosmology }\label{ap.b2}

We consider a flat FLRW Universe, whose line element is 
\begin{align}
ds^2&=-N^2(t)dt^2+a^2(t)d\omega^2, \label{metric}
 \end{align}
here $d\omega^2$ denotes the spatial spherical symmetry, the functions $a(t)$, $N(t)$ are the scale factor and the lapse, respectively. The action associated with $f_{nc}(R)$ gravity is
\begin{equation}
S=\int_{\mathcal{M}}{dx^4\sqrt{-g}f_{nc}(R)}+S_m,\label{acc}
\end{equation}
being $f_{nc}$ in the gravity sector, a Starobinsky-type function, 
\begin{equation}
f_{nc}(R)=a_{\eta,0}+a_{\eta,1}R+a_{\eta,2}R^2,\label{fun.frncc}
\end{equation}
where the subindex $\eta$ indicates the presence of the parameter of non-commutativity, and their explicit relation is given by (for details see Ref. \cite{fr})\footnote{Also, the manuscript shows that the scale factor satisfies 
$a\propto\Big[ \text{sech}\Big(\frac{3\sqrt{3}t}{2\sqrt{A}}\Big)\Big]^{\frac{1}{3}}$.}
\begin{align}
a_{\eta,0}&=\frac{2\kappa V_0e^{ 6\eta}}{e^{ 6\eta }+4},\nonumber\\
a_{\eta,1}&=\frac{\kappa V_0e^{ 3\eta }}{[e^{ 6\eta }+
4](6\times10^{-n_0})}\nonumber\\
&\quad+\Big[12e^{ 3\eta }+48e^{- 3\eta } \Big](6\times10^{-n_0})\ell^2,
\nonumber\\
a_{\eta,2}&=\Big[18e^{3\eta}+72e^{- 3\eta }\Big](6\times10^{-n_0}).
\label{fin.ncc}
\end{align}
mentioning that $\ell$ represents the extrinsic curvature associated to \eqref{metric} and $n_0>1$ a fixed appropriate integer. Now, varying \eqref{acc} with respect to the metric - introducing \eqref{fun.frncc} - we finally obtain the field equations. 
\begin{eqnarray}
f^{\prime}_{nc}(R)R_{\mu\nu}-\frac{1}{2}g_{\mu\nu}f_{nc}(R)+g_{\mu\nu}\square f^{\prime}_{nc}(R)-\nonumber\\
\nabla_{\mu}\nabla_{\nu} f^{\prime}_{nc}(R)=\kappa T_{\mu\nu}\label{f.e}
\end{eqnarray}
with $\square\equiv \nabla_{\mu}\nabla^{\nu}$ and $f^{\prime}_{nc}(R)\equiv \partial_{R}f_{nc}$.

%%%%%%%%%%%%%%%%%%%%%%%%%%%%%%%%%%%%%%%%%%%%%%%%%%%
\section{Mathematical Background}\label{ap.b}
%%%%%%%%%%%%%%%%%%%%%%%%%%%%%%%%%%%%%%%%%%%%%%%%%%%

\begin{lem}
Within a $f(R)$ gravity context, where $f$ is a Starobinsky-type function i.e. $f(R)=a_0+a_1R+a_2R^2$ 
together with the conditions shown in \cite{fr}, we have
\begin{itemize}
\item{Exists $t_0\in \mathbb{R}\setminus \{\pm \infty\}$ such that $\vert\dot{R} \vert<\epsilon$, for $\epsilon>0$ and $\vert t\vert>t_0 $.
}
\item{If $\beta_t=\text{Sech}^2(\mu t)$ and $I_n=(-\frac{t_0}{n},\frac{t_0}{n})$ then, $\vert\dot{\beta_t}  \vert<\epsilon$ \it{(a.e)} for $t\in I_1$
or the sequence\footnote{
$\mu$ is a number chosen appropriately, (a.e) means almost everywhere and the Lebesgue measure $m( \cdot)$ is considered.}
\begin{equation}
x_n(t)=\max_{t\in I_n}\vert\dot{\beta_t}\vert,\label{seq}
\end{equation}
converges to zero (in measure).}
\end{itemize}
\end{lem}
\begin{de}
Under the aforementioned scenario, the Ricci scalar satisfies $\vert\frac{R}{a^6}\vert<1$, and can be asserted that given $\rho,\delta>0$ there exists $t_1, t_2$ such that if $\vert t\vert>t_1$ we have $\vert\beta_t\vert<\rho$ and $\vert\dot{ \beta_t}\vert<\delta$ for $\vert t\vert>t_2$. We observe that $\vert R(t)\vert=\vert\beta_t\vert+\mathcal{O}(\rho)$, obtaining 
\begin{equation}
\frac{\vert R(t+\rho)\vert-\vert R(t)\vert}{\rho}=\frac{\vert\beta_{t+\rho}\vert-\vert\beta_t\vert}{\rho}
\end{equation}
or for small enough $\rho$ and $\vert t\vert>t_0=\max{\{t_1,t_2\}}$
\begin{equation}
\vert\dot{R}(t)\vert\approx\vert\dot{ \beta_t}\vert<\epsilon=\min{\{\rho,\delta\}}, \quad \vert t\vert>t_0.
\end{equation}

Now, for $t \in I=(-t_0,t_0)$
\begin{align}
\vert\dot{ \beta_t}\vert&<2\mu(e^{8\mu t_0}-1).
\end{align}
Since the maximum for $\vert\dot{ \beta_t}\vert$  can be found in $t\approx 1$ and for a small $ \delta$ we can find a rational $r_0$ in $(\mu t_0,  \mu t_0 + \delta$), getting the following inequality
\begin{equation}
\frac{1}{2\mu }\ln{\Big[  4^{-1}\text{sech}^2(\mu)\text{tanh}(\mu)+1 \Big] }\leq t_0<\frac{r_0}{\mu }, 
\end{equation}
where we can distinguish two cases:
\begin{equation}
\left\lbrace
\begin{array}{ll}
\frac{1}{8}\leq t_0<+\infty,\quad \text{if}\quad \mu \to 0\\\\
 0\leq t_0<\epsilon, \quad \text{for}\quad \mu \gg 1, \epsilon\downarrow 0. 
\end{array}
\right.
\end{equation}
First, if $\mu \gg 1$ then, for all $\epsilon>0$ we have 
\begin{equation}
m(I=\{t :  \vert\dot{ \beta_t}\vert>\rho \})=2\epsilon,\quad \text{for some}\quad \rho>0
\end{equation}
and $\vert\dot{\beta_t}\vert<\epsilon$ \it{(a.e)}.

Taking $0<\mu<1$ and $M_n=e^{ \frac{2\mu t_0}{n}}-1$ occurs $M_{n}\underset{n \to+\infty}{\longrightarrow} 0$ then, for $\epsilon>0$ 
exists $N_{ \epsilon}$ satisfying
\begin{align}
t_0&<\ln{ [  (\epsilon+1)^{ \frac{ n}{2\mu}} ]    },\quad n>N_{ \epsilon}.
\end{align}
Let \eqref{seq} and the set $Y_{n}=\{t\in I_n :  \vert x_n\vert\geq\rho\}$ for some $\rho>0$. If $\rho>2\mu(e^{2\mu t_0}-1)$, then $m(Y_{n}=\{\emptyset\})=0$, in other case, 
$m(Y_{n})\leq 2t_0$ and for $\epsilon=n^{-2}$ when $n>N_{ \epsilon}$
\begin{align}
m(Y_n)&\leq
\frac{ 1}{\vert\mu\vert}\Big\vert\sum_{i\geq 1}\frac{(-1)^{i+1}}{in^{2i-1}}\Big\vert.\label{serie}
\end{align}
We notice that, for the second term, the sequence of partial sums converges to zero
and the inequality \eqref{serie} can be written $m(Y_{n})\underset{n \to+\infty}{\longrightarrow} 0$, which means $x_n\overset{m}{\longrightarrow} 0$. $\blacksquare$
\end{de}
Additionally, to estimate $\tilde R$ we refer again to Ref. \cite{fr}, where the scale factor $a(t)\propto\Big[\text{sech}\Big(\frac{3\sqrt{3}t}{2\sqrt{A}}\Big)\Big]^{\frac{1}{3}}$ can be treated for the $f_{nc}$ cosmology scenario and the constant of proportionality is $C=\sqrt{A}e^{\frac{3\eta}{2}}[18\sqrt{a_{\eta 2}}]^{-1}$. Then, for a flat FLRW Universe we have
\begin{eqnarray}
\tilde{R}&=&\frac{R}{H^2_0}=\frac{6}{H^2_0}\Big(\frac{\ddot{a}}{a}+\frac{(\dot{a})^2}{a^2}\Big)\nonumber\\
&=&\frac{45}{2AH^2_0}\text{tanh}^2\Big(\frac{3\sqrt{3}t}{2\sqrt{A}}\Big)-\frac{9}{2AH^2_0},\nonumber\\
\end{eqnarray}
and when $\vert t \vert > 1$
\begin{eqnarray}
\tilde{R}
&=\frac{18}{AH^2_0}-\frac{90}{AH^2_0}e^{\frac{-3\sqrt{3}\vert t \vert}{\sqrt{A}}}+\mathcal{O}\Big(\frac{45}{2AH^2_0}e^{\frac{-3\sqrt{3}\vert t \vert}{\sqrt{A}}}\Big),\nonumber\\
\end{eqnarray}
obtaining
\begin{align}
\Big\vert \tilde{R}-\frac{18}{AH^2_0}+\frac{90}{AH^2_0}e^{\frac{-3\sqrt{3}\vert t \vert}{\sqrt{A}}}\Big\vert&\leq \frac{45M}{2AH^2_0}e^{\frac{-3\sqrt{3}\vert t \vert}{\sqrt{A}}}, \nonumber\\
\end{align}
with $M<+\infty$. Here, we impose 
\begin{equation}
    \frac{90}{AH^2_0}e^{\frac{-3\sqrt{3}\vert t \vert}{\sqrt{A}}}\leq \frac{10^{-4}}{H^2_0}\;\; {\rm and} \;\; \frac{45M}{2AH^2_0}e^{\frac{-3\sqrt{3}\vert t \vert}{\sqrt{A}}}\leq \frac{10^{-4}}{H^2_0},
\end{equation}
together with
\begin{equation}
\vert t \vert\geq \min \Bigg\{\frac{\sqrt{A}}{3\sqrt{3}}\ln{\Big[\frac{90\times10^{4}}{A}\Big]}, \frac{\sqrt{A}}{3\sqrt{3}}\ln{\Big[\frac{45M\times10^{4}}{2A}\Big]}\Bigg\},\nonumber
\end{equation}
finally getting
\begin{equation}
\Big\vert \tilde{R}-\frac{18}{AH^2_0} \Big\vert\leq\epsilon\to 0, \quad \text{or}\quad \tilde{R}\approx \frac{18}{AH^2_0}=\gamma_\eta.\label{A}
\end{equation}
For example, since $1\leq A\approx \frac{27V_0 10^{n_0}}{30}<+\infty$, taking $A=1$ we obtain \eqref{A}, for $\vert t \vert\geq 1.9$. Now, if $\vert t \vert \leq 1 $
\begin{align}
\tilde{R}&=\frac{18}{AH^2_0}\Big(\frac{135 t^2}{16A^2}-\frac{1}{4}\Big),\nonumber\\ 
\end{align}
for $\vert t \vert \to 0$, thus $\vert \tilde{R} \vert\approx\gamma_\eta$.
As a final comment, at the light of this lemma, is possible with \eqref{eq:gama} to read eq. \eqref{h2} 
(in measure) as 
\begin{align}
H^2&=\frac{\Big[2\kappa\rho_i+2f^{\prime}_{nc}(\bar{\gamma}_\eta)-f_{nc}(\bar{\gamma}_\eta)\Big]}{6f^{\prime}_{nc}(\bar{\gamma}_\eta)},\nonumber\\
2\dot{H}+3H^2&=-\frac{\Big[2\kappa \omega_i \rho_i +f_{nc}(\bar{\gamma}_\eta) \Big]}{2f^{\prime}_{nc}(\bar{\gamma}_\eta)}.\label{hae}
\end{align}
with $\bar{\gamma}_\eta=H^2_0\gamma_\eta$. The Friedmann constraint and relation \eqref{ratio} offers the next differential equation\footnote{Details are referred to Appendix \ref{ap.c}.} for $f_{nc}(\bar{\gamma}_\eta)$:
\begin{equation}
\frac{df _{nc}}{d\bar{\gamma}_\eta}-4f _{nc}=12H^2_0\left[(1\pm\nu)-\Omega_{0m}-\frac{3\Omega_{0r}}{4}\right], \label{ecdif}
\end{equation}
whose solution is
\begin{equation}
f_{nc}=c_1e^{4\bar{\gamma}_\eta}-3\Big[H^2_0(1\pm \nu)-\Omega_{m}-\frac{3}{4}\Omega_{r}\Big],\label{B.eq}
\end{equation}
getting for our purposes $f^{\prime} _{nc}=4c_1e^{4\bar{\gamma}_\eta}$,
where $e^{4\bar{\gamma}_\eta}\to 1$, in presence of 
\begin{equation}
\eta \in B(0,\zeta), \quad 
V_0\propto10^{-m}\quad  \text{and} \quad m<n_0. \label{gama1}
\end{equation}
Conditions \eqref{gama1} will be an important reference to study the dynamic in the $f_{nc}(R)$ context.  

%%%%%%%%%%%%%%%%%%%%%%%%%%%%%%%%%%%%%%%%%%%%%%%%%%%
\section{Justification of $\rho_m$ and $\rho_r$ positives and derivation of the dimensionless Friedmann equation} \label{ap.c}
%%%%%%%%%%%%%%%%%%%%%%%%%%%%%%%%%%%%%%%%%%%%%%%%%%%

Starting with equation (3) and making 
\begin{equation}
\beta^2\equiv f^\prime_{nc}\equiv\Big(2\sqrt{c_1}e^{2\bar{\gamma_\eta}}\Big)^2\equiv1\pm \nu,\label{nu}
\end{equation} for the positivity conditions
\begin{align}
\beta&<\sqrt{3}H_0\sqrt{\Omega_i(z+1)^{-3(\omega_i+1)}},
\end{align}
then, for $z\to 0$, we have 
\begin{align}
\beta&<\min\{\sqrt{3}H_0\sqrt{\Omega_m},\sqrt{3}H_0\sqrt{\Omega_r}   \}
=1.16, 
\end{align}
and $-1<\nu=(\beta+1)(\beta-1)<0.3456$. For the Ricci scalar
\begin{align}
10^{-n_0}&<R<\frac{1}{4}\ln{\Big[\frac{(1.16)^2}{4c_1}\Big]},
\end{align}
obtaining $c_1<0.3364$ and, when $c_1\approx 0.3364\times e^{-12.18H^2_0}$, we get 
$0<\tilde R<3.18$, which is a range in agreement with the values exposed in the manuscript.

 Now, considering \eqref{A} and introducing equation \eqref{eq:rho} in \eqref{hae}, we have
\begin{eqnarray}
&&3f^\prime_{nc}(\bar{\gamma_\eta})H^2=\Big(\rho_{0m}a^{-3}-f^\prime_{nc}(\bar{\gamma_\eta}) \Big)+\nonumber\\&&
\frac{3}{4}\Big(\rho_{0r}a^{-4}-f^\prime_{nc}(\bar{\gamma_\eta})\Big)+2f^\prime_{nc}(\bar{\gamma_\eta})\nonumber\\&&-f_{nc}(\bar{\gamma_\eta}),
\end{eqnarray}
after some manipulation, \textcolor{v}{it} is possible to find the expression
\begin{eqnarray}
&&f^\prime_{nc}(\bar{\gamma_\eta})H^2(z)=H^2_0\Big[\Omega_{m}(1+z)^{3}+\frac{3}{4}\Omega_{r}(1+z)^{4}\Big]+\nonumber\\&&
\frac{f^\prime_{nc}(\bar{\gamma_\eta})}{12}-
\frac{f_{nc}(\bar{\gamma_\eta})}{3}.
\end{eqnarray}
Imposing
\begin{align}
\frac{f^\prime_{nc}(\bar{\gamma_\eta})H^2(0)}{H^2_0}&=(1\pm \nu),\quad 0<\vert\nu\vert<1,
\end{align}
 to get a similar value for the Hubble parameter 
 at present, we obtain the differential equation \eqref{ecdif}, finding the solution
\eqref{B.eq}. Also, with \eqref{nu} follows
\begin{equation}
f_{nc}=\frac{\beta^2}{4}-3\Big[H^2_0(1\pm \nu)-\Omega_{m}-\frac{3}{4}\Omega_{r}\Big]\label{C},
\end{equation}
thus, we have
\begin{align}
&&f^\prime_{nc}(\bar{\gamma_\eta})E^2(z)&=\Omega_{m}(1+z)^{3}+\frac{3}{4}\Omega_{r}(1+z)^{4}+\frac{f^\prime_{nc}(\bar{\gamma_\eta})}{12H^2_0}\nonumber\\&&-\frac{f_{nc}(\bar{\gamma_\eta})}{3H^2_0}.
\end{align}
Then, substituting appropriately, we finally have the dimensionless Friedmann equation 
\begin{align}
\beta^2E^2(z)&=\Omega_{m}\Big[(1+z)^{3}-1\Big]+\frac{3}{4}\Omega_{r}\Big[(1+z)^{4}-1\Big]+\beta^2,
\end{align}
corresponding to Eq. \eqref{hf} in the main text.

\end{document}